\documentclass[preprint2]{aastex}

\shorttitle{The Sun's axisymmetric flux transport: 1996-2010}
\shortauthors{Hathaway \& Rightmire}

\begin{document}


\title{Variations in the axisymmetric transport of magnetic elements on the Sun: 1996-2010}
\author{David H. Hathaway}
\affil{NASA Marshall Space Flight Center, Huntsville, AL 35812 USA}
\email{david.hathaway@nasa.gov}

\author{Lisa Rightmire}
\affil{Department of Physics, The University of Alabama in Huntsville,
Huntsville, AL 35899 USA}
\email{lar0009@uah.edu}

\begin{abstract}
We measure the axisymmetric transport of magnetic flux on the Sun by cross-correlating narrow strips of data from line-of-sight magnetograms obtained at a 96-minute cadence by the MDI instrument on the ESA/NASA SOHO spacecraft and then averaging the flow measurements over each synodic rotation of the Sun. Our measurements indicate that the axisymmetric flows vary systematically over the solar cycle. The differential rotation is weaker at maximum than at minimum. The meridional flow is faster at minimum and slower at maximum. The meridional flow speed on the approach to the Cycle 23/24 minimum was substantially faster than it was at the Cycle 22/23 minimum. The average latitudinal profile is largely a simple sinusoid that extends to the poles and peaks at about $35\degr$ latitude. As the cycle progresses a pattern of in-flows toward the sunspot zones develops and moves equatorward in step with the sunspot zones. These in-flows are accompanied by the torsional oscillations. This association is consistent with the effects of the Coriolis force acting on the in-flows. The equatorward motions associated with these in-flows are identified as the source of the decrease in net poleward flow at cycle maxima. We also find polar counter-cells (equatorward flow at high latitudes) in the south from 1996 to 2000 and in the north from 2002 to 2010. We show that these measurements of the flows are not affected by the non-axisymmetric diffusive motions produced by supergranulation. 
\bigskip
\end{abstract}


\section{INTRODUCTION}

The structure and evolution of the magnetic field in the Sun's photosphere is believed to be produced by dynamo processes within the Sun \citep{Charbonneau05}. This structure and evolution must be faithfully reproduced in any viable dynamo model. Flux Transport Dynamo (FTD) models have recently been used to predict the strength of the next solar cycle \citep{Dikpati_etal06, Choudhuri_etal07}. In these FTD models the Sun's axisymmetric flows (differential rotaton and meridional flow) play key roles. The meridional circulation transports magnetic flux at the surface to the poles, builds up the polar fields, and sets the 11-year length of the solar cycle by its presumed slow equatorward return at the base of the convection zone. The differential rotation shears the poloidal magnetic field to produce strong toroidal fields that erupt through the photosphere in sunspots and active regions.

The structure and evolution of the photospheric magnetic field also serves as the inner boundary condition for all of space weather -- conditions on the Sun and in the space environment that can influence the performance and reliability of space-borne and ground-based technological systems. Surface Flux Transport (SFT) models have been used since 1984 \citep{DeVore_etal84} to evolve the surface field using the flux that erupts in active regions as a source term. This active region magnetic flux is then transported across the surface by meridional flow, differential rotation, and diffusion by supergranules -- nonaxisymmetric, cellular flows that evolve on a time scale of about 1-day. The magnetic field structure produced in SFT models has been used to model solar wind structures (wind speed and interplanetary magnetic field) for space weather forecasts \citep{ArgePizzo00} and to estimate the Sun's total irradiance since 1713 \citep{Wang_etal05} for Sun-Climate studies.

The strength, structure, and evolution of the meridional flow in particular is critically important in both FTD and SFT models. Unfortunately, the meridional flow is difficult to measure due to its weakness. Supergranules  have typical flow speeds of about 300 $\rm m\ s^{-1}$ and differential rotation has a typical velocity range of $\sim 200 \ \rm m\ s^{-1}$. Yet, the axisymmetric meridional flow has a top speed of only 10-20 $\rm m\ s^{-1}$.

The axisymmetric flows have been measured using a variety of techniques. Feature tracking
is amongst the simplest and oldest but gives different results depending on
the nature of the features themselves. Direct Doppler measurements can give the plasma
flow velocity in the photosphere but these measurements are subject to systematic
errors introduced by other solar processes and only provide the line-of-sight velocity
-- which, for the meridional flow, vanishes near the equator and limb.
Global helioseismology provides measurements of the differential rotation as a function of latitude, radius, and time.
Local helioseismology can provide measurements of the meridional flow as a function of
latitude, depth, and time using the methods of ring diagram
analysis or time-distance analysis.

Sunspots and sunspot groups were amongst the earliest features used to measure
the axisymmetric flows. \cite{Carrington59} measured the positions of sunspots on consecutive days and noted the presence of an equatorial prograde current and higher latitude retrograde flow. \cite{NewtonNunn51} measured the locations of recurrent sunspots groups on successive rotations as well as individual sunspots on consecutive days and found slightly different rotation profiles. \cite{Howard_etal84} made detailed measurements of individual sunspot positions recorded on photographic plates at Mount Wilson Observatory from 1921 to 1982. They found differential rotation with $\omega = 14.52 - 2.84 \sin^2 \lambda$ deg day$^{-1}$ (where $\lambda$ is the heliographic latitude) but noted that sunspot groups rotate more slowly than individual sunspots and large sunspots rotate more slowly than small sunspots.

Sunspots and sunspot groups can also be used to measure the meridional flow.
\cite{Tuominen42} used the latitudinal positions of recurrent sunspot groups and found
equatorward flow of $\sim 1 \ \rm m\ s^{-1}$ below $\sim 20\degr$ latitude and poleward
flow of similar strength at higher latitudes.
\cite{Ward73} used daily sunspot group positions to argue that there was no
meridional flow at the $1 \ \rm m\ s^{-1}$ level.
However, \cite{HowardGilman86} measured the latitudinal drift of individual sunspots and
found an equatorward flow of about $3 \ \rm m\ s^{-1}$ equatorward
of $\sim 25\degr$ with an even weaker poleward flow at higher latitudes.
An obvious drawback to tracking sunspots to measure the axisymmetric flows is the
limited latitudinal coverage (latitudes $< \sim 30\degr$) and the complete lack of
coverage at times near sunspot cycle minima.

Smaller magnetic features, although often concentrated in the active latitudes,
do cover the entire solar surface and are present even at sunspot cycle minima.
\cite{Komm_etal93A} masked out the active regions in high-resolution magnetograms
($2048 \times 2048$ pixel full-disk arrays) and cross-correlated
the remaining magnetic features with those seen the next day from 1975 to 1991 for several hundred magnetogram pairs.
They found differential rotation with $\omega = 14.43 - 1.77 \sin^2 \lambda - 2.58 \sin^4 \lambda$ deg day$^{-1}$ and noted that latitudinal profile was flatter at sunspot cycle maximum than at minimum.
\cite{Komm_etal93B} used the same technique to measure the meridional flow and found a poleward flow that varied with sinusoidally latitude, reaching a peak velocity of $\sim 13 \ \rm m\ s^{-1}$ at $39\degr$ latitude.
Furthermore, they found that the flow speed was slower at the sunspot cycle maximum than at minimum.
\cite{Meunier99} employed this technique (without masking the active regions) using magnetogram pairs from the MDI instrument \citep{Scherrer_etal95} on the ESA/NASA SOHO mission over the rising phase of sunspot cycle 23 from 1996 to 1998.
She found that the poleward meridional flow slowed in the presence of active regions.
In a recent paper \cite{HathawayRightmire10} did a similar analysis (with masking of the active regions) of MDI magnetograms over the time period from 1996 to 2009. They obtained measurements from over 60,000 image pairs separated by 8-hours.
They also found that the meridional flow was poleward (with a peak velocity of  $\sim 11 \ \rm m\ s^{-1}$ at $\sim 45\degr$ latitude) and was fast at cycle minimum but slow at cycle maximum.
In addition they noted that the speed of the meridional flow was substantially faster at the Cycle 23/24 minimum than at the Cycle 22/23 minimum.

Larger magnetic features, and associated structures, yield substantially different
results for the meridional flow. \cite{SnodgrassDailey96} cross-correlated Mt. Wilson
coarse array magnetograms ($34 \times 34$ pixel full-disk arrays) obtained 24-38
days (a solar rotation) apart and found poleward flow from $10\degr$ to $60\degr$
but equatorward flow at lower latitudes.
Their measurements extended from 1968 to 1992 -- covering three
sunspot cycle maxima and two minima. They also found a systematic dependence of the 
meridional flow pattern on the phase of the solar cycle. Out-flows from the
sunspot zones were observed to move toward the equator in step with the equatorward
movement of the sunspot zones themselves.
\cite{Latushko94} used the same low resolution data (after it was processed  to construct synoptic maps for each solar rotation) and also found out-flows from the sunspot zones.
\cite{Svanda_etal07} used a magnetic butterfly diagram constructed from synoptic maps of the magnetic field averaged over longitude for 180 equispaced zones in sine-latitude.
They measured the slope -- change in latitude vs. change in time -- of the magnetic
features and found a meridional flow with peak velocities of about $20 \ \rm m\ s^{-1}$ at the poleward limit ($\sim 45\degr$) of their measurements.

Here we measure the axisymmetric motions of the the small magnetic elements using the SOHO MDI data in which these elements are well resolved. These magnetic elements are presisely those whose transport is modeled in SFT models and in the surface transport of the FTD models.

\section{DATA PREPARATION}

High resolution full-disk images of the line-of-sight magnetic field have been obtained
at a 96-minute cadence since May 1996 by the SOHO MDI instrument.
These images were used in \cite{HathawayRightmire10} to find the variation in meridional
flow strength over solar cycle 23. They noted in that paper that the MDI imaging system
appears to be rotated by $\sim 0.21\degr$ counterclockwise with respect to the accepted position angle of the Sun's rotation axis. Furthermore, they found that the accepted position
of the Sun's rotation axis is in error by $\sim 0.08\degr$ as was noted previously by
\cite{Howard_etal84} and by \cite{BeckGiles05}. This small error introduced annual variations in the apparent cross-equatorial meridional flow. Here we account for those positional errors in mapping the full-disk magnetograms to heliographic coordinates by using modified values for the position angle and tilt of the Sun's rotation axis. In addition, while
reprocessing the data we found a significant reduction in the scatter of the
measurements if we took the MDI image origin to be at the bottom left corner of
the bottom left pixel -- not the center of the pixel as indicated in the MDI
documentation. Here we repeat the analyses in \cite{HathawayRightmire10} using these corrected magnetic maps and examine the variations in both the strength and structure of the axisymmetric flows.

\begin{figure}[ht]
\plotone{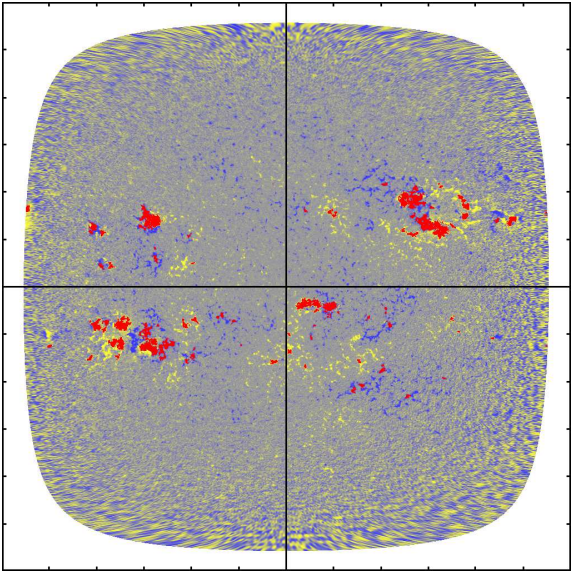}
\caption{
MDI magnetogram from 2001 June 5 04:48 UT mapped to heliographic coordinates.
Positive magnetic polarities are yellow, negative magnetic polarities are blue, and
masked areas are red. Tickmarks around the border are at $15\degr$ intervals in latitude and in longitude from the central meridian.
}
\end{figure}

Each full-disk magnetogram is mapped onto heliographic coordinates using bi-cubic interpolation onto a grid with 2048 by 1024 equispaced points in longitude and latitude for the entire surface of the Sun. This mapping gives a close match to the spatial resolution of the MDI instrument and makes longitudinal and latitudinal velocities linear functions of the displacements in the mapped coordinates.
The line-of-sight magnetic field is assumed to be largely radial so we divide the magnetic field strength at each image pixel by the cosine of the heliographic angle from disk center to minimize the apparent variations in field strength with longitude from the central meridian.
The magnetic fields in sunspots are intense enough to produce magnetic pressures similar to the plasma pressure (plasma $\beta \sim 1$). These intense magnetic field elements resist the near-surface plasma flow and have their own peculiar motions in longitude and latitude which vary depending on the size of the sunspot and age of the active region \citep{Howard_etal84}.
For this reason sunspots and their immediate surroundings are masked out. We found that this could be done quite effectively by identifying all mapped pixels with field strengths $\left|B\right| > 500$ G and all pixels within 5 mapped pixels of those points with $\left|B\right| > 100$ G as masked pixels.
An example of one of these mapped and masked magnetograms is shown in Fig. 1.

\section{ANALYSIS PROCEDURES}

The axisymmetric motions -- differential rotation and meridional flow -- of the magnetic
elements were determined by cross-correlating strips of pixels from pairs of mapped images
separated by 8 hours and finding the shift in longitude and latitude that gave the strongest correlation. (Results obtained with image pairs separated by 4.8 hours were substantially the same.) Each strip was 11 pixels ($\sim 2\degr$) high in latitude and 600 pixels
($\sim 105\degr$) long in longitude. The shift in longitude and latitude producing the
strongest correlation was calculated to a fraction of a pixel by fitting parabolas in
longitude and latitude through the correlation coefficient peaks. This process was performed
at 860 latitude positions from $75\degr$S to $75\degr$N for typically about 400 image
pairs over each 27-day rotation of the Sun. In all we obtained measurements from over 60,000 magnetogram pairs.

The average and the standard deviation of the differential rotation and meridional
flow velocities were calculated at each latitude for each solar rotation of 27.25 days.
The differential rotation and meridional flow profiles for each rotation were
fit with fourth order polynomials in $\sin \lambda$, where $\lambda$ is the heliographic latitude. Errors in the fit coefficients were estimated using a Monte Carlo
method with random variations at each latitude characterized by the standard deviations
from the measurements. These polynomial coefficients were also recast in terms of associated Legendre polynomials of the first order. The Legendre polynomial coefficients are better suited
for studies of time variations based on the orthogonality of the polynomials themselves \citep{Snodgrass84}.

The latitudinal profiles of differential rotation and meridional flow as measured with these data and this method represent the actual axisymmetric motions of the magnetic elements. Since the magnetic elements are fully resolved in these data the effects of supergranule diffusion are seen as random motions of the magnetic elements and these random motions do not introduce any systematic errors in our measurements as will be shown in Section 7.
Profiles were obtained for 178 rotations of the Sun from June 1996 to September 2010 with a gap from June 1998 to February 1999 when radio contact with SOHO was lost and not fully recovered.

\section{AVERAGE FLOW PROFILES}

The average differential rotation profile from the entire dataset is shown in Fig. 2.
The velocities are taken relative to the Carrington frame of reference which has a sidereal
rotation rate of $14.184\rm{\ deg\ day}^{-1}$. The average differential rotation profile is
well represented by just the three terms with symmetry across the equator --

\begin{equation}
v_\phi(\lambda) = (a + b \sin^2 \lambda + c \sin^4 \lambda) \cos \lambda
\end{equation}

\noindent with

\begin{equation}
a = 35.6 \pm 0.1 \rm{\ m\ s}^{-1}
\end{equation}

\begin{equation}
b = -208.6 \pm 1.1 \rm{\ m\ s}^{-1}
\end{equation}

\begin{equation}
c = -420.6 \pm 1.6 \rm{\ m\ s}^{-1}
\end{equation}

\noindent This gives an angular rotation rate profile with

\begin{equation}
\omega(\lambda) = A + B \sin^2 \lambda + C \sin^4 \lambda
\end{equation}

\noindent with

\begin{equation}
A = 14.437 \pm 0.001 \rm{\ deg\ day}^{-1}
\end{equation}

\begin{equation}
B = -1.48 \pm 0.01 \rm{\ deg\ day}^{-1}
\end{equation}

\begin{equation}
C = -2.99 \pm 0.01 \rm{\ deg\ day}^{-1}
\end{equation}

\noindent where coefficient $A$ includes the Carrington rotation rate.

This angular rotation rate is nearly identical to that found by \cite{Komm_etal93A} for the time interval 1975-1991 using similar data and methods.
We do find a slight north-south asymmetry as seen in Fig. 2 by the deviation of the measured profile from the symmetric profile given by the dashed line. The differential rotation was slightly weaker in the south than in the north. We also note the flattening of the profile at the equator with a slight ($\sim 1 \rm{\ m\ s}^{-1}$) but significant dip from $\pm 5\degr$ to the equator. A similar ``dimple'' at the equator was seen previously in direct Doppler data by \cite{Howard_etal80} and in magnetic element motions by \cite{Snodgrass83}.

\begin{figure}[ht]
\plotone{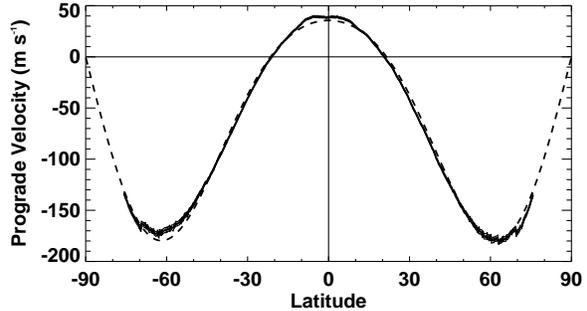}
\caption{
The average differential rotation profile with the $2\sigma$ error range for the
time interval 1996-2010. The symmetric profile given by Eqns. 1-4 is shown with the
dashed line.
}
\end{figure}

The average meridional flow profile for the entire dataset is shown in Fig. 3.
Although the average meridional flow profile does display substantial north-south
asymmetry, the profile is well represented with just the two anti-symmetric terms --

\begin{equation}
v_\lambda(\lambda) = (d \sin \lambda + e \sin^3 \lambda) \cos \lambda
\end{equation}

\noindent with 

\begin{equation}
d = 29.7 \pm 0.3 \rm{\ m\ s}^{-1}
\end{equation}

\begin{equation}
e = -17.7 \pm 0.7 \rm{\ m\ s}^{-1}
\end{equation}

\noindent This gives a peak poleward meridional flow velocity of $11.2 \rm{\ m\ s}^{-1}$
at a latitude of $35.2\degr$. This is somewhat slower than the meridional flow found by \cite{Komm_etal93B} for the time interval 1975 to 1991 but with a peak at nearly the same latitude. Our average meridional flow profile shows substantially different flows in the north and in the south. The flow velocity is faster in the south and peaks at a higher latitude than in the north. The flow in the north appears to nearly vanish at the extreme northern limit ($75\degr$) of our measurements while the flow in the south is still poleward with a speed of about $5\rm{\ m\ s}^{-1}$ at the southern limit.

\begin{figure}[ht]
\plotone{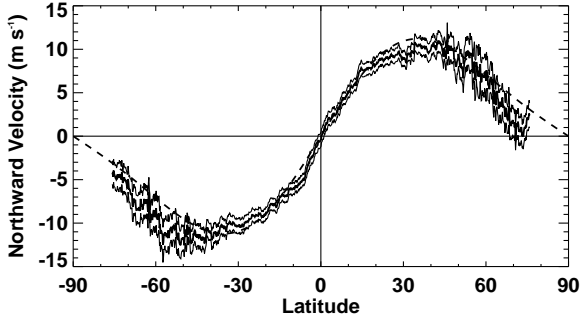}
\caption{
The average meridional flow profile with $2\sigma$ error range for the
time interval 1996-2010. The anti-symmetric profile given by Eqns. 9-11 is shown with the
dashed line. This profile shows substantially different flow in the north and south.
}
\end{figure}

\section{VARIATIONS IN FLOW SPEED}

Variations in the amplitudes of the axisymmetric flow components were examined
by plotting the rotation-by-rotation histories of the Legendre polynomial coefficients.
The Legendre polynomials were normalized so that their maximum values were either
1.0 or -1.0. The coefficients that multiply them then give the peak velocity for that
component. The normalized polynomials we used are

\begin{equation}
P_1^1(\lambda) = \cos \lambda
\end{equation}

\begin{equation}
P_2^1(\lambda) = 2 \sin \lambda \cos \lambda
\end{equation}

\begin{equation}
P_3^1(\lambda) = \sqrt{135 \over 256} (5 \sin^2 \lambda - 1) \cos \lambda
\end{equation}

\begin{equation}
P_4^1(\lambda) = 0.947 (7 \sin^3 \lambda - 3 \sin \lambda) \cos \lambda
\end{equation}

\begin{equation}
P_5^1(\lambda) = 0.583 (21 \sin^4 \lambda - 14 \sin^2 \lambda + 1) \cos \lambda
\end{equation}

\begin{figure}[ht]
\plotone{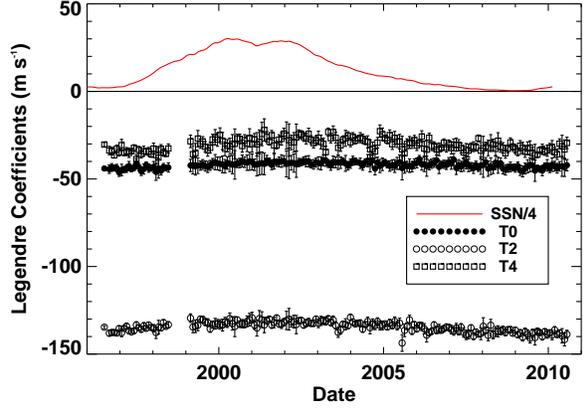}
\caption{
The differential rotation associated Legendre polynomial coefficients (with $2\sigma$ error bars) for the time interval 1996-2010.
The coefficient T0 multiplies $P_1^1$, the polynomial of zeroth order in $\sin \lambda$.  
The coefficient T2 multiplies $P_3^1$, the polynomial of second order in $\sin \lambda$.  
The coefficient T4 multiplies $P_5^1$, the polynomial of fourth order in $\sin \lambda$.
The smoothed sunspot number divided by 4 is shown in red for reference. The differential rotation is slightly weaker (flatter) at sunspot cycle maximum.  
}
\end{figure}

The Legendre coefficient histories for the differential rotation are shown in Fig. 4 along with the smoothed sunspot number for reference to the phase of the sunspot cycle.
The three symmetric components ($P_1^1$, $P_3^1$, and $P_5^1$) dominate so we only show the three associated coefficient histories.
These three coefficients show only a slight variation over the sunspot cycle with the
amplitudes being smaller (less negative -- weaker differential rotation) at sunspot cycle maximum ($\sim 2002$). This ``more rigid'' differential rotation at sunspot cycle maximum was previously noted by \cite{Komm_etal93A}.

\begin{figure}[ht]
\plotone{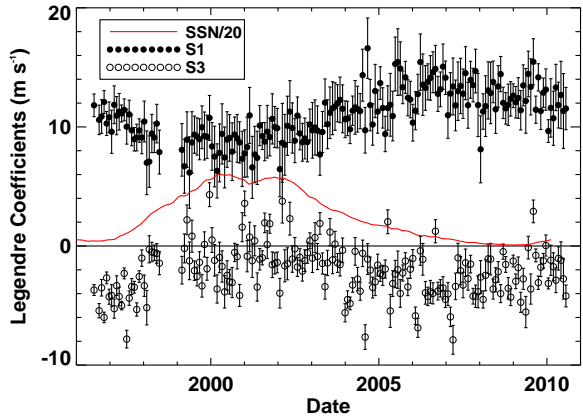}
\caption{
The meridional flow Legendre polynomial coefficients (with $2\sigma$ error bars)
for the time interval 1996-2010.
The coefficient S1 multiplies $P_2^1$, the polynomial of first order in $\sin \lambda$.  
The coefficient S3 multiplies $P_4^1$, the polynomial of third order in $\sin \lambda$.  
The smoothed sunspot number divided by 20 is shown in red. The meridional flow is slower at sunspot cycle maximum but was even faster at Cycle 23/24 minimum in 2008 than at Cycle 22/23 minimum in 1996. 
}
\end{figure}

The Legendre coefficient histories for the meridional flow are shown in
Fig. 5 along with the smoothed sunspot number.
The two anti-symmetric components ($P_2^1$, and $P_4^1$) dominate so we only show the two associated coefficient histories.
These two coefficients show substantial variations over the sunspot cycle with the
amplitudes being smaller at sunspot cycle maximum.
\cite{Komm_etal93B} found similar behavior for the time period 1978-1990.

In addition to this systematic trend over the sunspot cycle (fast at minimum and slow at maximum) we find a secular variation in which the meridional flow speed was substantially ($\sim 20\%$) faster at the
Cycle 23/24 minimum in 2008 than at the Cycle 22/23 minimum in 1996.
As in \cite{HathawayRightmire10} we note that the meridional flow speed was faster for the entire interval from 2004 on, than it was at the cycle minimum in 1996. This increase in meridional flow speed would explain the weak polar fields that were produced during that time period in the SFT models of \cite{SchrijverLiu08} and \cite{Wang_etal09}.

\section{VARIATIONS IN STRUCTURE}

The variations in flow speed shown in the last section are produced by and accompanied
by variations in flow structure. Our analyses produce latitudinal profiles of the differential rotation and the meridional flow for each individual solar rotation from June 1996 to September 2010. These profiles were obtained at 860 latitude positions between $\pm 75\degr$.
For further analysis we smoothed these profiles with a tapered Gaussian having a FWHM of 6 latitude points ($\sim 1\degr$), resampled at intervals of $1\degr$ in latitude, and produced images of these latitudinally smoothed profiles and of the differences between each such profile and the average symmetrized profiles. Little, if any, variation can be seen in the full differential rotation profile history. However, the meridional flow profile history shows substantial variation as shown in Fig. 6.

\begin{figure}[ht]
\plotone{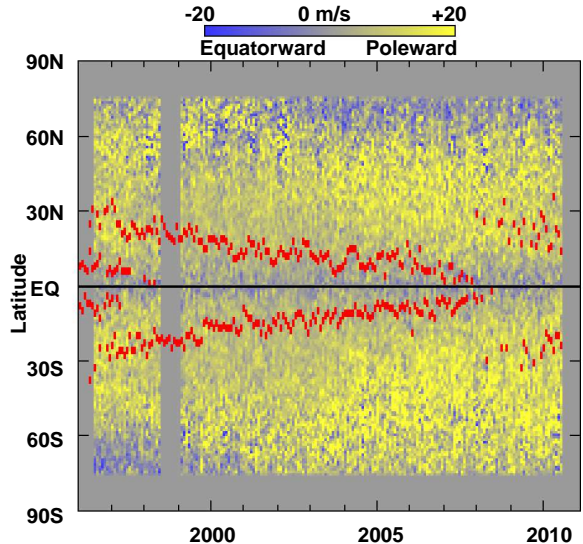}
\caption{
The meridional flow profiles for individual solar rotations from 1996-2010.
Poleward flow is indicated by shades of yellow. Equatorward flow is indicated by shades of blue. The latitudinal centroid of the sunspot area in each hemisphere for each rotation is shown in red. The weakening of the meridional flow in the active latitudes near sunspot cycle maximum is evident as are polar counter-cells (equatorward flow) in the south from 1996 to 2000 and in the north from 2002 to 2010.
}
\end{figure}

The structure of the meridional flow changes substantially over the time period represented in Fig. 6. The weakening of the poleward meridional flow at sunspot cycle maximum (1999-2003) is evident in the muted colors surrounding the sunspot zones. The strengthing of the meridional flow on the approach to Cycle 23/24 minimum in late 2008 is evident in the intensified colors at most latitudes after 2004.

Fig. 6 also reveals the existence of counter-cells (equatorward flow). One is found in the south extending equatorward to about $60\degr$S at the start of the dataset in May of 1996 but that boundary moves poleward of our $75\degr$ limit by mid-2000. A similar counter-cell is seen forming in the north in 2002 as it dips below $75\degr$N and remains in evidence to the end of the dataset in 2010. This long-lasting northern counter-cell is clearly the primary source of the north-south asymmetry seen in the average meridional flow profile (Fig. 3) and may be associated with the asymmetry in the differential rotation (Fig. 2). The fact that it maintains its existence for more than half of the time available in this dataset leaves its imprint on the average meridional flow profile in the form of the rapid drop in poleward flow in the north to near zero at $75\degr$N latitude.

\begin{figure}[ht]
\plotone{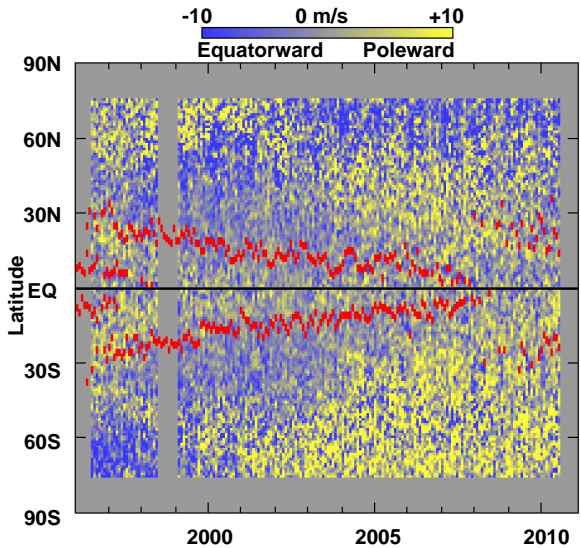}
\caption{
The differences between the meridional flow profiles for individual solar rotations and the average, symmetric profile from 1996-2010.
Poleward flow (relative to the average profile) is indicated by shades of yellow. Equatorward flow is indicated by shades of blue. The latitudinal centroid of the sunspot area in each hemisphere for each rotation is shown in red. The system of in-flows toward the sunspot zones is evident as poleward flow on the equatorward sides of the sunspot zones and equatorward flow on the poleward sides.
}
\end{figure}

Additional details concerning the structural changes in the axisymmetric flows are seen when the average symmetric flow profiles are subtracted from the profiles for each individual rotation. These differences from the average for the meridional flow are shown in Fig. 7.
The two counter-cells are more obvious here. In addition, these difference profiles show a system of in-flows (relative to the average meridional flow) toward the sunspot zones with poleward (yellow) flows on the equatorward sides and equatorward (blue) flows on the poleward sides. This suggests that the slowdown in the poleward meridional flow seen at sunspot cycle maxima is produced by the growing strength and latitudinal extent of these in-flows.

The presence of these in-flows was nonetheless somewhat surprising. \cite{SnodgrassDailey96} found \emph{out-flows} from the active latitudes with their low-resolution magnetic data. \cite{ChouDai01} and \cite{Beck_etal02} also found out-flows from the active latitudes using time-distance helioseismology. However, \cite{GonzalezHernandez_etal10} found clear evidence for in-flows much like what we see in Fig. 7 using ring-diagram helioseismology and the structural changes seen in the magnetic element motions by \cite{Meunier99} also support the presence of these in-flows. 

\begin{figure}[ht]
\plotone{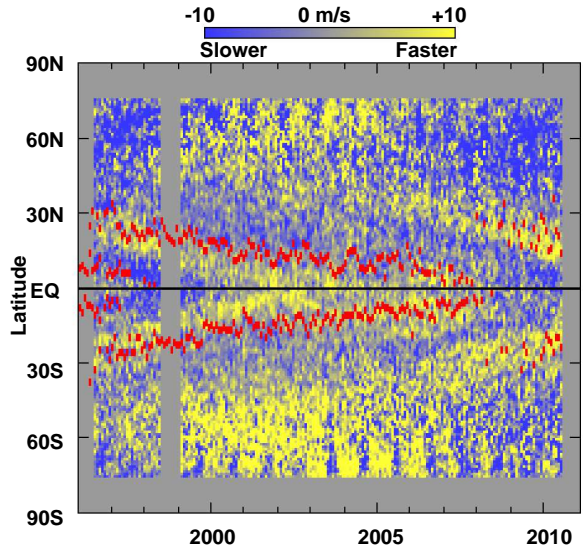}
\caption{
The differences between the differential rotation profiles for individual solar rotations and the average, symmetric profile from 1996-2010.
Faster (prograde relative to the average profile) flow is indicated by shades of yellow. Slower (retrograde) flow is indicated by shades of blue. The latitudinal centroid of the sunspot area in each hemisphere for each rotation is shown in red. The torsional oscillations are evident as faster flow on the equatorward sides of the sunspot zones and slower flow on the poleward sides. 
}
\end{figure}

The in-flows toward the sunspot zones are accompanied by the torsional oscillations -- variations in the differential rotation seen as faster rotation on the equatorward sides of the sunspot zones and slower rotation on the poleward sides \citep{HowardLaBonte80}. This is shown in Fig. 8 by the differences in the differential rotation profiles from the average symmetrized differential rotation profile. (Note that there are instrumental artifacts at the highest latitudes as evident by the annual variations in flow speed with faster flow near the poles in the hemisphere tilted toward the observer. These artifacts may be due to an elliptical distortion of the MDI image as reported by \cite{Korzennik_etal04}. However, our efforts to include this distortion with either the angle they reported or the angle given in the MDI documentation did not improve the results.)

These variations in the differential rotation are consistent with the effect of the Coriolis force on the in-flows and the counter-cells. Material moving equatorward from the higher latitudes will spin-down and give slower flows on the poleward sides of the sunspot zones while material moving poleward from the equator will spin-up and give faster flows on the equatorward sides. This scenerio was suggested by \cite{Spruit03} as a response to cooling in the sunspot zones by excess thermal emission from faculae. Earlier, \cite{Snodgrass87} had suggested that in-flows and the torsional oscillations were part of a system of azimuthal convection-rolls which migrate equatorward during each sunspot cycle. These convection-rolls should have out-flows at some undetermined depth below the surface -- a possible source of the out-flows seen in some of the helioseismology studies. The Coroilis force acting on the long-lasting northern counter-cell should slowdown the rotation at the affected latitudes. This may be the source of the north-south asymmetry in the average differential rotation profile (Fig. 2).

\section{EFFECTS OF DIFFUSION ON FLOW MEASUREMENTS}

The magnetic elements under study here are also subject to a diffusion-like random walk by the nonaxisymmetric cellular flows -- supergranules in particular \citep{Leighton64}. This random walk transports the weak magnetic elements in both longitude and latitude and leads to the formation of large unipolar areas from the preceding and following magnetic flux in active regions \citep{Smithson73}. This random walk might contribute to the meridional flow we measure due to resultant changes in the magnetic pattern. In SFT models \citep{DeVore_etal84, vanBallegooijen_etal98, Wang_etal02, Wang_etal05, Wang_etal09, SchrijverLiu08} this process is represented by a diffusivity coupled with the Laplacian of the magnetic field. We would expect that this might produce a meridional flow signal in the form of out-flows from the sunspot zones where the magnetic field is concentrated. Although what we observe is actually in-flows toward the sunspot zones, the effects of diffusion might nonetheless alter the structure and evolution of the meridional flow we measure. Given this caveat, we undertook an investigation of the effects of supergranule diffusion on our measurements.

\cite{Hathaway_etal10} have recently produced a model of the photospheric flows which includes the cellular flows, supergranules in particular, observed with the SOHO MDI instrument. The cellular flows in this model have velocity spectra, lifetimes, and motions that match those seen in the MDI data itself. We have taken the vector velocities from this model and used them to transport magnetic elements whose initial spatial distribution was taken from an MDI synoptic magnetic map. We then used our analysis procedures to measure the axisymmetric flows. We isolated the effects of diffusion by only including the evolving cellular flows. We do not include the axisymmetric meridional flow or differential rotation and the cellular flow pattern itself does not participate in any axisymmetric meridional flow or differential rotation.

The cellular flow simulation produced vector velocities on a heliographic grid with 4096 by 1500 equispaced points in longitude and latitude from an evolving velocity spectrum that extended to spherical wavenumbers of 1500 (supergranules have spherical wavenumbers of $\sim 100$). The initial magnetic field distribution was taken from an MDI synoptic magnetic chart for Carrington rotation 2000 (mid-2003 -- just after the peak of the sunspot cycle). Our magnetic flux transport simulation was calculated on a grid the same size as our mapped magnetograms. At each pixel in our simulated magnetic map we introduced a number of 1000 G magnetic elements with filling factors of 5\% until the average field strength in that pixel equaled the observed field strength (a single element in a pixel would produce a field strength of 50 G). This process required some 120,000 magnetic elements. These elements were then transported explicitly by the velocity field from the cellular flow simulation in 15-minute time steps for 10 days. 

\begin{figure}[ht]
\plotone{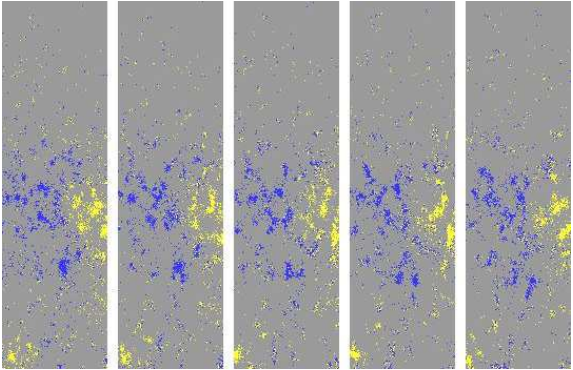}
\caption{
Simulated magnetic map regions at 1-day intervals. These regions were extracted from the full simulated magnetic maps at the start of days 1-5 from an area bordered by the equator, $60\degr$N, and longitudes $109\degr$ and $126\degr$. The evolving magnetic network is evident in the changing magnetic structures.
}
\end{figure}

Examples from the simulated magnetic maps are shown in Fig. 9. The magnetic elements are transported to the borders of the cells and then continue to move as the cells themselves evolve. (This was shown in previous simulations by \cite{Simon_etal01}.) The magnetic elements retain their identities throughout the simulation and do not interact with each other. If opposite polarities occupy a pixel they do cancel each other in terms of the mapped magnetic field strength but they continue to retain their identities and move with the simulated flow.

These magnetic maps were processed with the same analysis procedures used with the MDI magnetic maps by selecting a ``central meridian'' longitude and correlating strips of pixels with those from a map 8-hours later. This was done for a series of cental meridians at 1-hour intervals over the 10 simulated days. This resulted in 559 measurements of the axisymmetric flows covering the full range of longitudes and the full 10 days. Fig. 10 shows the meridional flow measured from these magnetic maps. The results have similar noise levels to single rotation averages from MDI but show no evidence of any systematic meridional flow.

\begin{figure}[ht]
\plotone{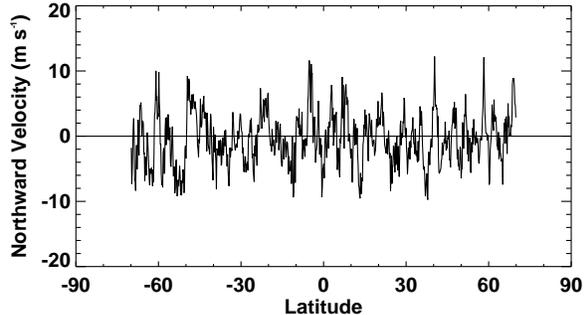}
\caption{
Meridional flow profile measured from magnetic features subjected to random walk by non-axisymmetric cellular flows. Our meridional flow measurements do not include any systematic errors due to these random (and spatially resolved) motions.
}
\end{figure}

\section{CONCLUSIONS}

We have measured the axisymmetic motions of magnetic elements on the Sun by cross-correlating strips of data from magnetic maps acquired at 96-minute cadence by the MDI instrument on SOHO. Our measurements cover each rotation of the Sun from June 1996 to September 2010 with the exception 8 rotations when the data were unavailable. Although we exclude the magnetic elements in sunspots themselves, the magnetic elements we track are in fact those whose poleward motions produce the Sun's polar fields in SFT models \citep{DeVore_etal84, vanBallegooijen_etal98, Wang_etal02, Wang_etal05, Wang_etal09, SchrijverLiu08} and in FTD models \citep{Dikpati_etal06, Choudhuri_etal07}. With these data these magnetic elements are well resolved and the random motions due to supergranules appear as just that -- random motions that do not alter our measurements of the axisymmetric flows.

The differential rotation we measure agrees well with previous measurements using similar data and methods \citep{Komm_etal93A}. Although the average differential rotation profile is slightly asymmetric this asymmetry may be specific to the time period and the presence of the meridional flow counter-cell in the north. The torsional oscillation signal (Fig. 8) compares well with the near surface pattern from helioseismology \citep{Howe_etal09} and does not require averaging the two hemispheres together.

The meridional flow we measure also agrees well with previous measurements using similar data and methods \citep{Komm_etal93B, Meunier99} but with interesting differences and more detail. The average meridional flow speed we found from 1996 to 2010 was somewhat slower than found by \cite{Komm_etal93B} from 1978 to 1991. We both find that the flow is faster at cycle minima and slower at maxima. Here we find that this slow-down can be attributed to a system of in-flows toward the sunspot zones which, when superimposed on the average meridional flow profile, lowers the peak flow velocity at cycle maxima \citep{Meunier99}. Our slower average meridional flow speed is somewhat surprising since our data included two (fast) minima and one maximum while the \cite{Komm_etal93B} data included two (slow) maxima and one minimum.

An important difference for understanding the long, drawn-out, and low Cycle 23/24 minimum is the faster meridional flow after 2004 compared to the flow at the Cycle 22/23 minimum in 1996. This faster meridional flow produces weaker polar fields in the SFT models of \cite{SchrijverLiu08} and \cite{Wang_etal09}. Weaker polar fields produce weak following cycles which typically have long, low minima \citep{Hathaway10}.

In spite of this agreement, our average meridional flow profile is problematic for the SFT models. All of the SFT modeling groups use meridional flow profiles which peak at low latitudes or do not extend poleward of $75\degr$. Comparisons between our symmetrized profile and those used in three SFT calculations \citep{vanBallegooijen_etal98, Wang_etal09, SchrijverTitle01} are shown in Fig 11.

\begin{figure}[ht]
\plotone{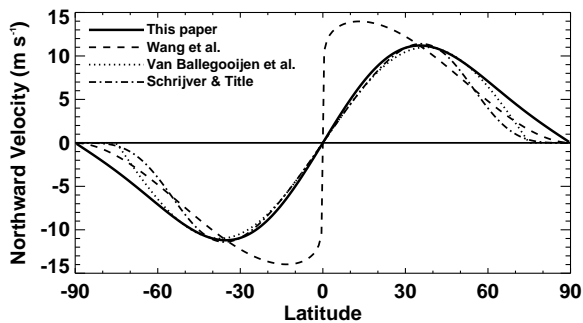}
\caption{
Symmetrized meridional flow profile from this paper (solid line) plotted with meridional flow profiles use in the Surface Flux Transport models of \cite{Wang_etal09} (dashed line) \cite{vanBallegooijen_etal98}  (dotted line) and \cite{SchrijverTitle01} (dot-dashed line).
}
\end{figure}

All three SFT profiles fall below our measured profile at the higher latitudes -- above $30\degr$ for \cite{Wang_etal09}, $45\degr$ for \cite{SchrijverTitle01}, and $60\degr$ for \cite{vanBallegooijen_etal98}. Using our average meridional flow profile in these models without compensating processes leads to polar fields substantially stronger than those observed. Compensating processes might include the counter-cells along with the north-south asymmetry or neglected physical processes -- for example radial diffusion suggested by \cite{Baumann_etal06}.

The nearly 20\% change in meridional flow speed from Cycle 22/23 minimum in 1996 to Cycle 23/24 minimum  in 2008 is problematic for the FTD models. \cite{DikpatiCharbonneau99} showed that with their FTD model increasing the surface meridional flow speed from $2\rm{\ m\ s}^{-1}$ to $20\rm{\ m\ s}^{-1}$ changed the surface polar field strength from 130G to 350G while changing the cycle period from 77 years to 11 years. The faster meridional flow in this model should have produced a shorter cycle with stronger polar fields. Yet, observations reveal a very long cycle with much weaker polar fields.

We have shown that our data, with its high spatial resolution and rapid cadence, fully resolve the magnetic element motions produced by supergranule ``diffusion'' and thus yield measurements of the meridional flow without any systematic errors due to that diffusion. \cite{Komm_etal93B} used data with similar spatial resolution but lower cadence (daily rather than hourly) and found similar results. However, \cite{SnodgrassDailey96} and \cite{Latushko94} used data with much lower spatial resolution and much longer time-lags (monthly) and found significant differences. These low spatial resolution data do not resolve the individual magnetic elements. They image the emsemble magnetic patches whose motions \emph{do} include the effects of diffusion. We suspect that the magnetic pattern diffusion gave the equatorial flows at low latitudes measured by \cite{SnodgrassDailey96} and the out-flows from the sunspot zones seen by \cite{SnodgrassDailey96} and \cite{Latushko94}, and more rapid high-latitude flow seen by \cite{Svanda_etal07}.

Comparisons of our measurements with those from other data types (direct Doppler velocities, sunspot motions, and helioseismology) are subject to problems associated with the characteristic depth of the measurements. The Sun has a surface shear layer produced largely by the granule and supergranule flows which tend to conserve angular momentum \citep{FoukalJokipii75} -- slowing down the rotation of the surface layers and speeding up the rotation down to depths of about 35 Mm. This inward increase in rotation rate should be accompanied by an inward decrease in the meridional flow speed \citep{Hathaway82} -- a feature noted by \cite{Hathaway_etal10} in the meridional motion of supergranules. This is consistent with the slower rotation rate and faster meridional flow seen in direct Doppler measurements representative of the photosphere \citep{Ulrich_etal88, Ulrich10} assuming that the magnetic elements are rooted in somewhat deeper layers. Sunspots should be rooted even deeper yet and sunspots show rotation rates which are even more rapid \citep{Ward66, Howard_etal84} and meridional motions that are vanishingly small \citep{Ward73} or equatorward \citep{Tuominen42, HowardGilman86}. While helioseismology studies indicate both out-flows \citep{ChouDai01, Beck_etal02} and in-flows \citep{GonzalezHernandez_etal10}, this may be due to differences in both the methods used and the associated depths of the measurements. Helioseismology does provide supporting evidence for the variations in meridional flow speed over the sunspot cycle \citep{BasuAntia03, GonzalezHernandez_etal10}.

Our observations of in-flows toward the sunspot zones may help us understand the origins of the torsional oscillations. The strength and structure of these in-flows are good matches to the flows predicted in the model of \cite{Spruit03}. However, helioseismology indicates that the torsional oscillations may originate well below the surface at high latitudes \citep{BasuAntia03} and thus may not be forced by the effects of localized surface cooling.

Finally, we reitterate our point that the magnetic elements whose motions we study are precisely those elements whose transport is modeled in SFT models and at the surface in FTD models. Both SFT and FTD models must employ the measured axisymmetric transport of those magnetic elements to conform with observations.

\acknowledgements
DH would like to thank NASA for its support of this research through a grant
from the Heliophysics Causes and Consequences of the Minimum of Solar Cycle 23/24 Program to NASA Marshall Space Flight Center. LR would like to thank NASA for its support through an EPSCoR grant to Dr. Gary P. Zank through The University of Alabama in Huntsville.
SOHO, is a project of international cooperation between ESA and NASA.

\end{document}